\documentstyle[12pt,epsf]{article}

\def \dfrac #1#2 {\displaystyle\frac{#1}{#2}}
\def\be{\begin{eqnarray}}
\def\ee{\end{eqnarray}}
\def\bq{\begin{equation}}
\def\eq{\end{equation}}
\def\ben{\begin{enumerate}}\def\een{\end{enumerate}}

\def\prl {Phys. Rev. Lett. }\def\pr{Phys. Rev. }
\def\np{Nucl. Phys. }\def\pl{Phys. Lett. }

\def\roughly#1{\mathrel{\raise.3ex\hbox{$#1$\kern-.75em%
\lower1ex\hbox{$\sim$}}}}

\begin{document}
\begin{titlepage}
%\hfill {\large Preliminary}
\hfill \today

\vspace{.3cm}
%\hfill {\large FTUV 97-29; IFIC 97-29}
\vspace{.2cm}
\begin{center}
\ \\
{\Large \bf A quark model analysis
\\
\vspace{.1cm}
of the transversity distribution:
Next to leading order evolution$\dagger$}
\ \\
\ \\
\vspace{1.0cm}
{Sergio Scopetta and Vicente Vento$^{(a)}$}
\vskip 0.5cm
{\it Departament de Fisica Te\`orica}

{\it Universitat de Val\`encia}

{\it 46100 Burjassot (Val\`encia), Spain}

            and

{\it (a) Institut de F\'{\i}sica Corpuscular}

{\it Consejo Superior de Investigaciones Cient\'{\i}ficas}
\end{center}
\vskip 1.0cm
\centerline{\bf Abstract}
\vskip 0.4cm
The feasibility of measuring chiral-odd parton distribution functions in
polarized Drell-Yan and semi-inclusive experiments has renewed theoretical
interest in their study . Not long ago we presented an analysis 
of the transversity distribution functions using models of hadron 
structure and Leading Order evolution. In the meantime Next to Leading Order 
evolution equations for $h_1$ are also available. We here present a 
comparison
between the LO and the NLO results which confirms our previous conclusions.

\vskip 1cm
\leftline{Pacs: 12.39-x, 13.60.Hb, 13.88+e}
\leftline{Keywords:  hadrons, partons, transversity, evolution.}
\vspace{.7cm}

{\tt
\leftline {scopetta@titan.ific.uv.es}
\leftline{vicente.vento@uv.es}}
\vspace{0.7cm}
\noindent{\small$\dagger$Supported in part by DGICYT-PB94-0080, 
DGICYT-PB95-0134
and TMR programme of the European Commisison ERB FMRX-CT96-008}
\end{titlepage}

In  recent work we studied the transversity distributions 
using models of hadron structure \cite{sv97} . Our scheme consists
of two well defined procedures. Firstly we calculate the leading
twist contributions in a low energy model of hadron structure. These results 
are associated with a low $Q^2$, the so called hadronic scale \cite{jr80}. 
Secondly we move our results from the hadronic scale to the
experimental conditions using perturbative $QCD$ evolution \cite{tr97}. 
In our previous work we used Leading Order evolution, since this was the 
only 
one available for the transversity distributions at that time. 
Within a short 
period of time three calculations have appeared which allow the 
generalization 
of the evolution procedure to Next to Leading Order in this case 
\cite{km97,wv97,hk97}.  It can be shown that the last two calculations 
agree 
with each other but not with the first one. 
Therefore we shall assume that the 
correct result is that of refs.\cite{wv97,hk97} and proceed to analyze if 
the conclusions of our paper are mantained to this order.

We  discuss the results for the proton in two models
\begin{itemize}
\item[i)] The non relativistic model of Isgur-Karl \cite{ik78};
\item[ii)] The relativistic MIT bag model \cite{ja74}.
\end{itemize}

In both cases we will use the corresponding support correction as defined in
\cite{tr97} and \cite{jr80}, respectively. In Fig. 1 we show the results
corresponding to $g_1$ and $h_1$ for the IK model, corresponding
to the pure valence quark hadronic scenario of ref.\cite{tr97}, characterized
by a very low hadronic scale 
\footnote{As discussed in ref \cite{tr97}, the 
hadronic scale for LO is $\mu_0^2 = 0.079$ GeV$^2$, while for the
NLO case is $\mu_0^2 = 0.094$ GeV$^2$, parameters which are fixed
by assuming that the valence quarks carry all the momentum.}.
The initial data are evolved to 10 GeV$^2$, by using in the Next to Leading
evolution the $\overline{MS}$ factorization 
scheme of refs. \cite{wv97,hk97} and the evolution parameters 
of ref. \cite{tr97}.

The figure shows that the diverse evolution properties of these two structure
functions lead to a large difference between the two initially identical
functions. The difference occurs at small $x$ and has been noted also 
by other
authors \cite{bc96,ba97}. We compare in this case the LO and 
NLO evolutions of $g_1$ and $h_1$ and conclude that the 
results are quite stable
with respect to the perturbative evolution. Similar results hold for 
the other model.

We next turn another important result of our previous paper, namely
the analysis of Soffer's inequalities. As has been pointed out by
Vogelsang \cite{wv97}, in NLO the factorization scheme plays an important
role in defining them. All the results shown will be in the $ \overline{MS}$
scheme.

Both models verify the primitive 
Soffer inequality \cite{so95}, i.e., 
\bq
2|h_1^q(x,Q^2)| \leq g_1^q(x,Q^2) + f_1^q(x,Q^2),
\eq 
not only at the hadronic scale, but also as we evolve the distribution 
function
towards the experimental regime. However Soffer argued \cite{so95} that his 
positivity bound could be used combined with data to limit the validity of 
models. In particular by imposing the simple relation
\bq
\Delta u(x) = u(x) - d(x)
\eq
proposed in \cite{bs95} and which is well supported by the data, 
it is possible
to use the positivity bound  to obtain the allowed range of values
for $h^u_1$, namely
\bq
u(x) - d(x)\ge |h^u_1(x)|~.
\label{soffer1}
\eq
The MIT bag model  fails the bound for large values of $x$ \cite{so95}.
We show in Fig. 2 the comparison of the experimental constraint at 4 GeV$^2$
with the Isgur-Karl and MIT bag calculations. 
In the figure the allowed region
is described by taking the lefthand side of eq. (\ref{soffer1}) 
from the data.
The remaining curves represent the righthand side of the equation which
we have calculated from the models. It is clear from the figure that,
at the hadronic scale, neither fulfils the constraint.
However the above inequalities are valid at the scale of the data, then
we must compare only after evolving the model calculations at NLO order  
from the hadronic scale to that of the data (4 GeV$^2$).
As the figure shows, consistency is achieved after this procedure. 
This result does not imply that the conventional models of hadron structure
taken as a description of the physics at the hadronic scale are 
quantitatively
succesful in explaining the deep inelastic data. 
As stated in previous analysis
\cite{tr97}, these models give a qualitative description, which we have 
confirmed for the Soffer inequalities. However in order to obtain a 
quantitative description additional ingredients have to be added 
(see \cite{sv97,sv197}
for a further discussion).

The use of models of hadron structure to describe 
the deep inelastic properties
of the proton and neutron has proven successful for the chiral-even twist two
structure functions \cite{tr97} (and references therein). Several authors 
have generalized  the analysis  to the transversity functions 
\cite{bc96,st93,jj94,jm97}. 
Since these have not been measured, this analysis 
has the added value of prediction. We have completed the spectrum 
of possible 
calculations by including that of a well established not relativistic model, 
with a fine tuned technique for constructing the structure functions and 
performing the RGE evolution \cite{sv97}. Moreover we have returned to the 
highly succesful field theoretic approach of the MIT 
bag model and reanalized 
some of the features questioning  its validity, i.e., the highly discussed
Soffer inequalities \cite{sv97}.
In this note we have simply checked the stability of our 
results by performing
evolution to the Next to Leading Order and therefore we have corroborated the
conclusions of our previous work.

\section*{Acknowledgements}
\indent\indent

We have mantained iluminating discussions with Marco Traini and acknowledge
some tuition on the evolution code. One of us (S.S.) thanks
A. Drago, R. Jacob and P.J. Mulders for fruitful discussions.

\newpage
\centerline{\bf \Large Captions}
\hfill\break
\hfill\break
\hfill\break
{\bf Figure 1}:
 We show the trasversity function $h_1(x,\mu_0^2)$
(continuous line) which coincides with the spin distribution function
$ g_1(x,\mu_0^2)$  for the Isgur-Karl model \cite{ik78} at the hadronic
scale;
% $\mu_0^2 = 0.079$ GeV$^2$;
their evolved (LO and NLO)  distributions $h_1(x,Q^2)$ (dotts and 
long-dashed) 
and $g_1(x,Q^2)$ (dot-dashed and dashed)) at $Q^2 = 10$ GeV$^2$ are  shown. 
The (NLO) evolutions have been performed in the $ \overline{MS} $ 
factorization 
scheme \cite{wv97,hk97} and using the parameters taken from ref. \cite{tr97}
(cf. Fig. 1 (a) in ref. \cite{sv97}).
\hfill\break
\hfill\break

{\bf Figure 2}:
The allowed region of Soffer determined from the experimental
data at 4 GeV$^2$, corresponds to the region inside the continuous line. 
Fig. a): the dashed line corresponds to the pure Isgur-Karl model calculation
\cite{ik78}; the dot-dashed line represents the evolved IK solution from an
hadronic scale of 0.094 GeV$^2$, while the dotted line corresponds to the 
evolved solution of the IK model supplemented by gluons from an hadronic 
scale of 0.3 GeV$^2$. Fig b): the dashed line corresponds to
the pure MIT calculation; the dotted line assumes a hadronic scale of 0.75
GeV$^2$, while the dot-dashed line one of 0.094 GeV$^2$.
%, consistent with the
 Stratmann analysis \cite{st93}. 
The evolution has been carried out to Next to Leading Order in the 
$\overline{MS}$ factorization scheme and 
the evolution parameters have been taken from ref. \cite{tr97}
(cf. Fig. 4 in ref. \cite{sv97}).

\newpage

\begin{figure}[h]
\vspace{10cm}
\includegraphics{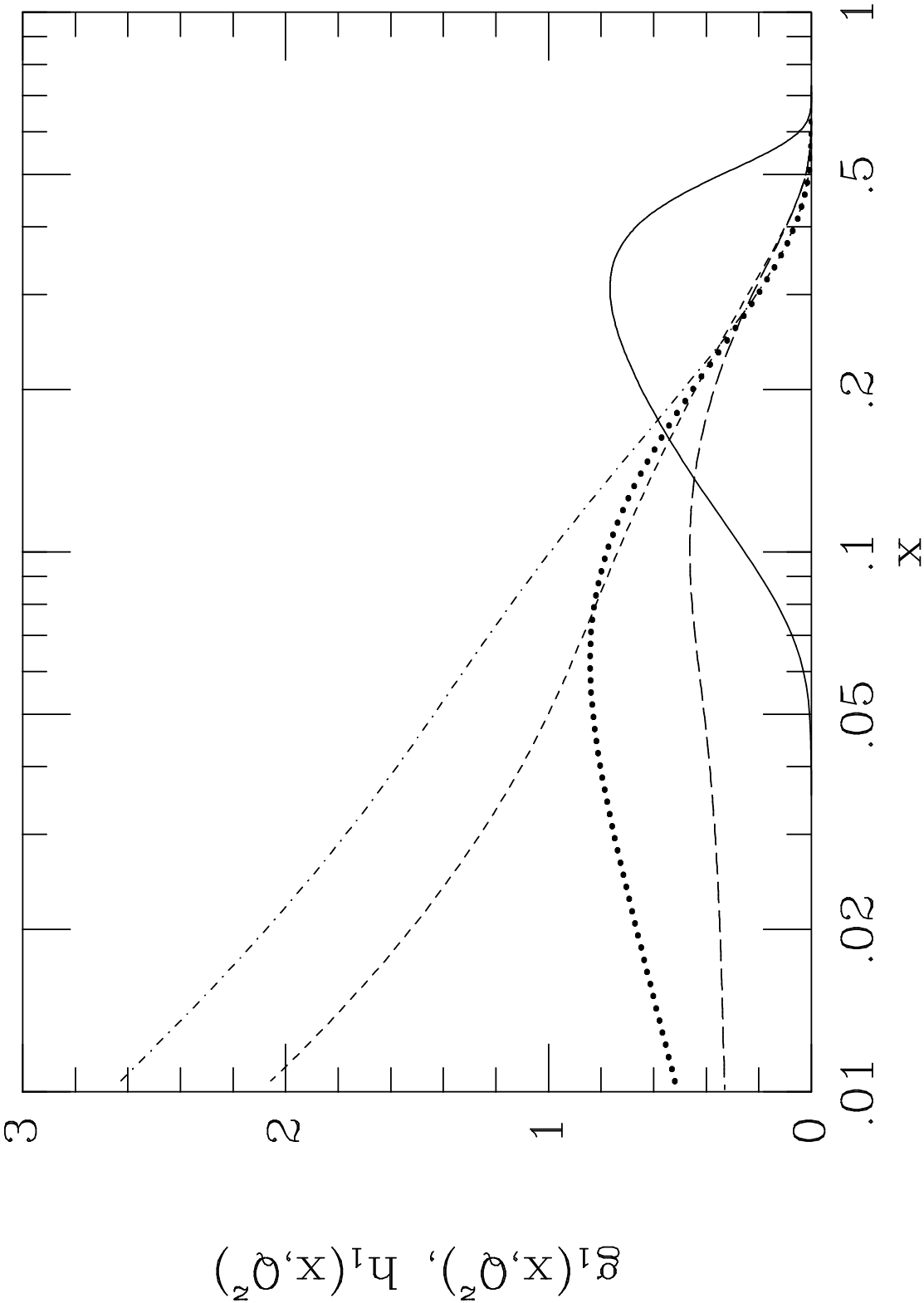}
%\special{psfileik2.ps hoffset240 voffset185 hscale 31 vscale=
%33 angle270}
\end{figure}
\vspace{2cm}
%\centerline{\large S. Scopetta and V. Vento}
\vspace{1cm}
\centerline{\bf \large FIGURE 1}

\newpage

\begin{figure}[h]
\vspace{10cm}
\includegraphics{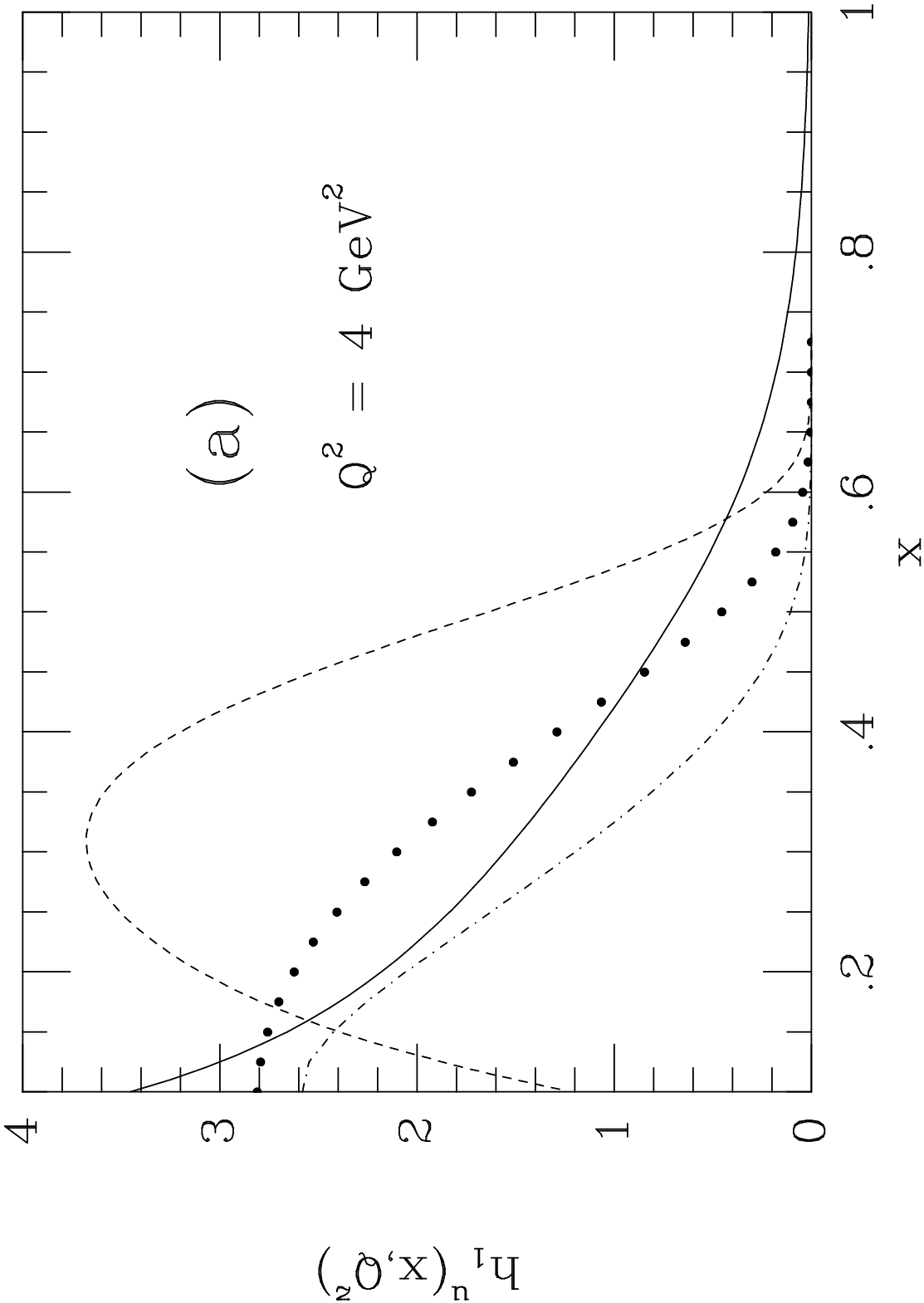}
\includegraphics{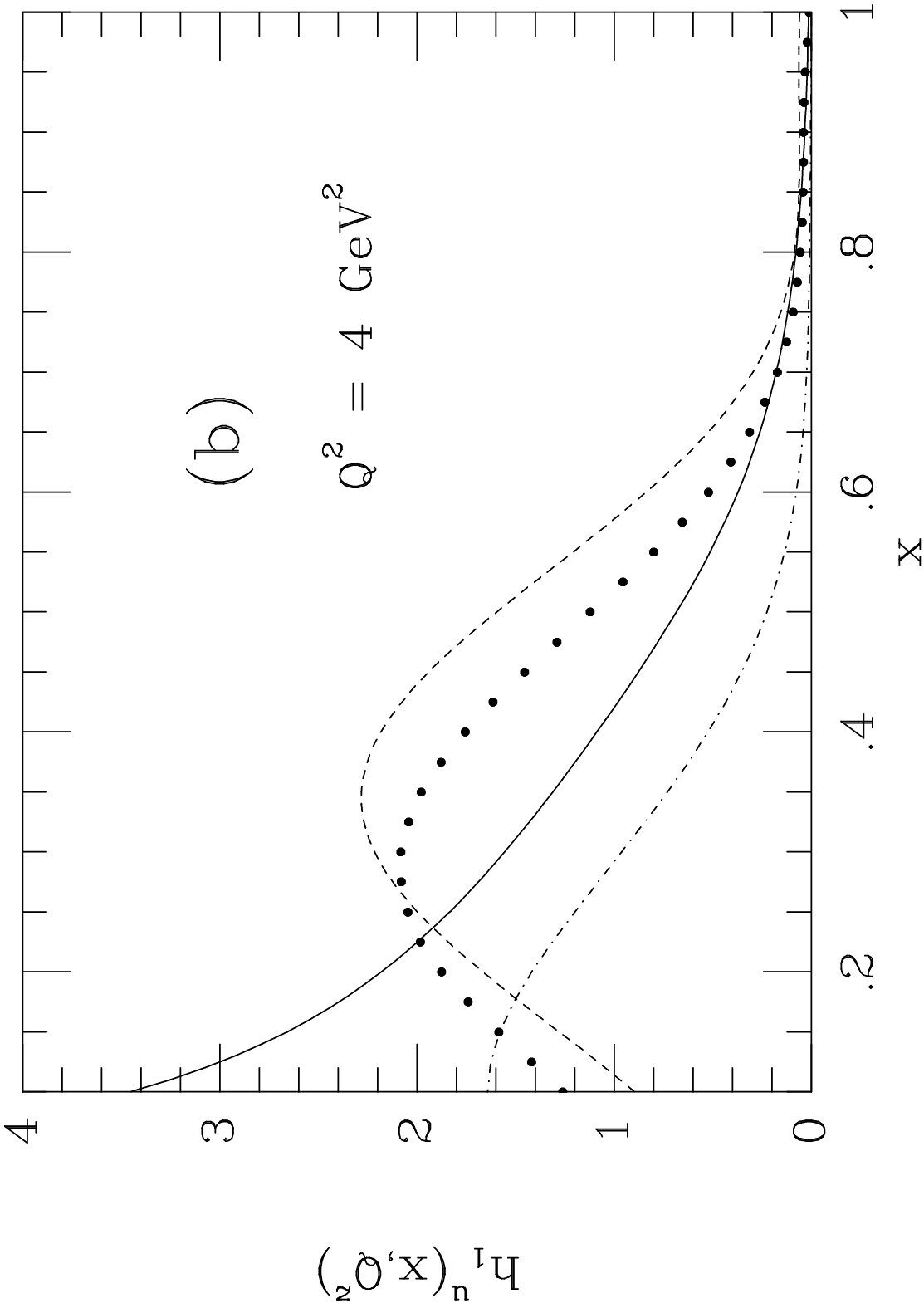}
\end{figure}
\vspace{2cm}
%\centerline{\large S. Scopetta and V. Vento}
\vspace{1cm}
\centerline{\bf \large FIGURE 2}

\end{document}